# Real-time Spectroscopy with Sub-GHz Resolution using Amplified Dispersive Fourier Transformation

Jason Chou, Daniel R. Solli, and Bahram Jalali

*Department of Electrical Engineering*

*University of California, Los Angeles, California 90095*

Abstract

Dispersive Fourier transformation is a powerful technique in which spectral information is mapped into the time domain using chromatic dispersion. It replaces a spectrometer with an electronic digitizer, and enables real-time spectroscopy. The fundamental problem in this technique is the trade-off between the detection sensitivity and spectral resolution, a limitation set by the digitizer's bandwidth. This predicament is caused by the power loss associated with optical dispersion. We overcome this limitation using Raman amplified spectrum-to-time transformation. An extraordinary loss-less -11.76 ns/nm dispersive device is used to demonstrate single-shot gas absorption spectroscopy with 950 MHz resolution—a record in real-time spectroscopy.

Real-time spectroscopy is useful for studying the elusive dynamics found in chemical reactions, biological systems, and other high-speed applications. Unfortunately, conventional spectrometers typically employ moving components, which are not conducive to fast single-shot measurements. Chirped Wavelength Encoding and Electronic Time-domain Sampling (CWEETS) overcomes this obstacle, offering unprecedented spectral acquisition rates by eliminating the use of a conventional spectrometer [1-4]. Originating from the work on photonic time-stretch analog-to-digital conversion [5], the method employs a spectrum-to-time transformation to physically Fourier transform the signal, allowing the spectrum to be measured directly in the time domain. Its advantages are two-fold. First, its pulsed nature provides extremely rapid spectral acquisition capability that can span hundreds of nanometers in a single shot. Second, its time-domain operation uses a single photodetector and a high-speed electronic digitizer to acquire extremely high resolution spectra in real-time. CWEETS has been applied to absorption spectroscopy for binary chemical detection as well as the study of combustion dynamics [2-4].

Until recently, practical limitations have significantly restricted the spectral resolution of CWEETS well below the performance of conventional spectrometers. The loss of the dispersive medium, the essential component that performs spectrum-to-time mapping, limits the detection sensitivity as well as the spectral resolution. The latter can be understood by recognizing that spectral resolution is fixed by the temporal resolution of the electrical detection system. Stated differently, the electrical bandwidth of the digitizer limits the spectral resolution, a relation given by $\Delta\lambda = (0.35 \times D \times f_{dig})^{-1}$, where $\Delta\lambda$ is the spectral resolution, $D$ is the total group velocity dispersion and $f_{dig}$ is the input bandwidth of the real-time digitizer. The fundamental trade-off in CWEETS is self evident: the product of $\Delta\lambda \times D$ is fixed by the bandwidth of the digitizer. To increase the optical resolution (i.e., to reduce $\Delta\lambda$) one is forced to increase the total group velocity dispersion, $D$. But this comes at the expense of increased optical loss and reduced detection sensitivity. This dispersion-loss trade-off is a fundamental connection described by the Kramers-Kronig relations [6], which indicate that broadband dispersive elements are inevitably lossy. The loss in the dispersive element is, therefore, the central problem in CWEETS.

Sampling oscilloscopes offer larger electrical bandwidth compared with real-time digitizers (roughly a factor of five, for current technology), and hence, a better optical resolution. However, this speed comes at the expense of single-shot operation because sampling scopes do not capture the signal in real time.

In this letter, we demonstrate sub-GHz spectral resolution in single-shot absorption spectroscopy through the use of a Raman-amplified spectrum-to-time transformation. Amplified CWEETS has been recently applied to real-time stimulated Raman spectroscopy [7]. Here, we apply an amplified dispersive element to perform real-time high-resolution absorption spectroscopy of narrow gas lines. The amplification produces a virtually "loss-less" dispersive element with an extraordinarily large -11.76 ns/nm of dispersion, and allows us to reach a record spectral resolution of 950 MHz for a single-shot measurement. With this highly dispersive loss-less element, we can also oversample the absorption spectrum of a narrow gas line to boost the signal-to-noise ratio of the single-shot measurement while maintaining high spectral resolution.

Distributed Raman amplification (DRA) within DCF has been deployed in long-haul optical communication links [8]. It is well known that DCF exhibits a Raman gain peak 10 times greater than that of normal single mode fiber, owing to its small effective core area [8]. Serendipitously, DCF is also the preferred medium for CWEETS because of its relatively high dispersion-to-loss ratio as well as its very low group delay distortion (in contrast to chirped fiber Bragg gratings (CFBGs)).

A major advantage of using DRA, in contrast to discrete optical amplifiers, is its ability to maintain a constant signal power level throughout the dispersive element. This important property maximizes the signal-to-noise-and-distortion ratio (SNDR) by keeping the signal power away from low power (noisy) and high power (nonlinear) regimes, evoked by traditional discrete amplifiers, e.g., erbium-doped fiber amplifiers (EDFAs) [8]. Furthermore, Raman amplification has another significant advantage: in optical fiber, an amorphous medium, it is naturally broadband. The gain spectrum can be further broadened through multiple pump lasers, and extremely broadband gain spectra, important for spectroscopy, can be realized using incoherent pump sources [7]. Raman-amplified dispersive elements also eliminate the need for high-power supercontinuum sources that have been required for high-resolution CWEETS in the past. High energy optical pulses are undesirable because they have the potential to cause changes or even permanent damage to the sample, as well as unwanted nonlinearities in the dispersive element. Even in situations where these issues are not significant, DRA can still be employed to reduce the power demands on the supercontinuum source.

Figure 1 illustrates the experimental setup of a Raman-amplified CWEETS system. A broadband pulse, with a 600 fs pulse width, -15 dBm/nm power spectral density, and 1.25 MHz repetition rate, is chirped by a -11.76 ns/nm dispersive element. The signal evolution inset shows the non-uniform spectral profile of the source and its transformation into a time-domain waveform. The Raman-amplified dispersive element uses an optical circulator and fiber-coupled Faraday mirror to produce a double-pass configuration—providing a two-fold improvement in GVD over the usual single-pass approach. An optical bandpass filter carves a 15 nm band centered at 1590 nm from the pulse. The DCF is pumped by four 350 mW continuous-wave lasers positioned at regular length intervals located at 1470, 1480, 1490, and 1480 nm. These pump wavelengths are chosen to produce a uniform Raman gain profile across the optical filter bandwidth with ~1 dB variation. The Raman gain bandwidth (15 nm in the present work) is arbitrarily determined by the number of pump lasers available. Distributed Raman amplification can operate in any spectral region where suitable pumps and fibers are available; using standard fibers, Raman gain bandwidths of up to 136 nm have been readily demonstrated [8], and the gain bandwidth can be located anywhere in the transparency window between roughly 1300 and 1650 nm typically used for fiber optic communications. In this paper, a counter-propagating pumping scheme is chosen to further improve the SNR and reduce the nonlinear distortion penalty [8]. The net gain of the Raman amplified DCF element is set to zero.

The signal evolution inset of Fig. 1 shows a cartoon of the rotational-vibrational absorption lines encoded in the time domain, before and after normalization of the pulse envelope. A 10 GHz photodiode and 20 GSa/s (4 GHz analog bandwidth) real-time oscilloscope are used to capture and digitize the waveform. The receiver's rise time value (93 ps) is used to define the minimum temporal resolution. Given the total GVD, the single-shot spectral resolution is estimated to be 950 MHz (8 pm or 0.03 cm$^{-1}$). We point out that a state-of-the-art 18 GHz, 60 GS/s real-time oscilloscope (LeCroy SDA18000) may be readily used to achieve a spectral resolution of 1.6 pm in a single shot with the same optical setup. The update time for measuring real-time dynamics is determined by the repetition rate of the source laser, currently 800 ns. For extremely fast dynamic processes, time-division and wavelength-division multiplexing techniques for high speed optical communication links can be applied to further reduce the update time.

The definition of spectral resolution is a qualitative assessment that depends on the application of interest. The rise time of the electronic receiver represents a convenient and representative figure of merit for estimating the spectral resolution of CWEETS. Proper interpretation of this number is subject to the particular application. For example, distinguishing two closely spaced absorption peaks could result in a slightly better spectral resolution than what is calculated using rise-time. On the other hand, measuring the width of an ultrafine absorption line may produce a slightly larger value.

The digital data acquisition format of CWEETS is ideal for adopting powerful digital signal processing techniques, which can be used to improve the signal-to-noise ratio (SNR). To examine the waveform properties of the normalized gas signal, we calculate a fast Fourier transform (FFT), as shown in Figure 2. The signal bandwidth below 1.5 GHz, falling easily within the available 4 GHz oscilloscope bandwidth, indicates that the measured gas signal is digitized beyond the minimum rate required by the Nyquist theorem, a condition known as oversampling. In this situation, the surplus of samples may be averaged in order to enhance the SNR and maintain single-shot acquisition [9], at the cost of reduced spectral resolution. Theoretically, each factor of two increases in the sampling rate beyond the Nyquist criterion can decrease the SNR by 3 dB [9].

To demonstrate the performance of CWEETS, we measure the absorption spectrum of a gas cell at room temperature containing carbon monoxide ($^{12}C^{16}O$) (NIST 2514) with a pressure of 133 kPa and path length of 80 cm. Figure 3a shows raw data from a single-shot measurement with an 8 pm (950 MHz) spectral resolution. Next, a 1.5 GHz RF digital filter is applied to delete high frequency noise. Using the rise-time estimation, the digital filter reduces the spectral resolution to 47 pm; however, a clear SNR enhancement is obtained (cf. Figure 3b). The reduced resolution is still sufficient for resolving the ~50 pm linewidth of the carbon monoxide sample in a single shot. In other applications, such as the study of biochemical dynamics, processes often occur on millisecond timescales—in these cases, shot-by-shot averaging can be employed to enhance the SNR. In contrast to oversampling, shot-by-shot averaging does not reduce the spectral resolution. Figure 3c shows the spectrum after 100 consecutive shots are averaged in time, corresponding to an update time of 80 microseconds.

To further demonstrate the ultra-high resolution capabilities of Raman-amplified CWEETS, we illustrate single-line absorption measurements with and without SNR enhancement. The dotted curve shown in Figure 4 illustrates a single-shot high-resolution measurement at 950 MHz (8 pm), without SNR enhancement. Although distortion due to noise is evident, the 50 pm linewidth of carbon monoxide may be accurately resolved without additional processing. In order to enhance the SNR in a single shot, oversampling may be utilized. As depicted by the dashed line, this method reduces spectral resolution to 47 pm, but provides a noticeable improvement in signal quality. For applications requiring high spectral resolution without rapid refresh rates, the SNR may be increased further by averaging 100 consecutive shots, shown in the solid line.

In conclusion, we demonstrate the application of Raman amplification to real-time absorption spectroscopy. We overcome a practical limitation in the spectral resolution of spectrum-to-time mapped spectroscopy by achieving extraordinarily high GVD values with zero net loss. This work demonstrates a single-shot spectrometer with sub-GHz optical spectral resolution. In addition, we show how oversampling can be used to enhance the SNR of a single-shot acquisition while maintaining high spectral resolution.

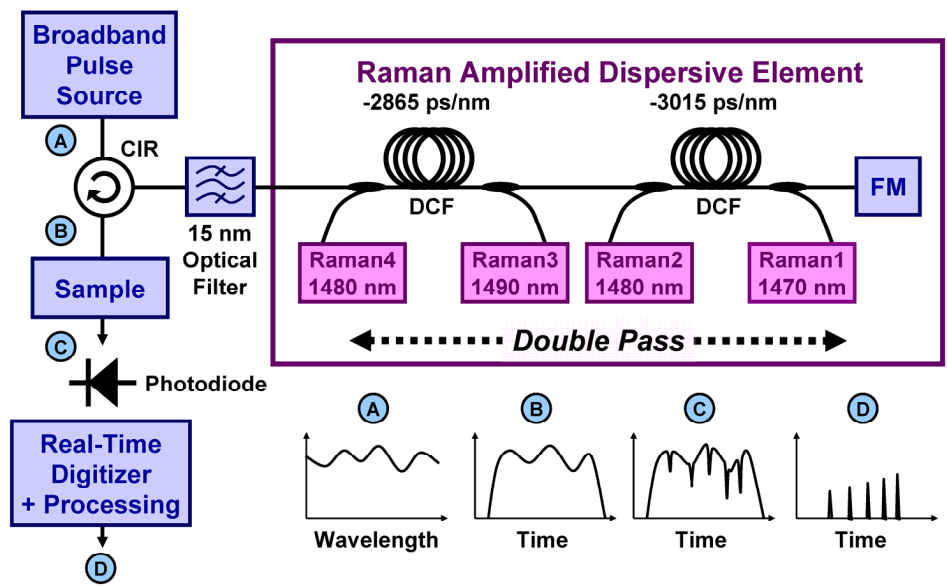

Figure 1: Experimental block diagram of the CWEETS spectrometer with 950 MHz frequency resolution. Absorption lines of a gas sample are encoded onto a chirped optical pulse and digitized by a real-time electronic oscilloscope. A -11.76 ns/nm "lossless" dispersive element is achieved using Raman amplification. CIR: circulator. FM: Faraday mirror.

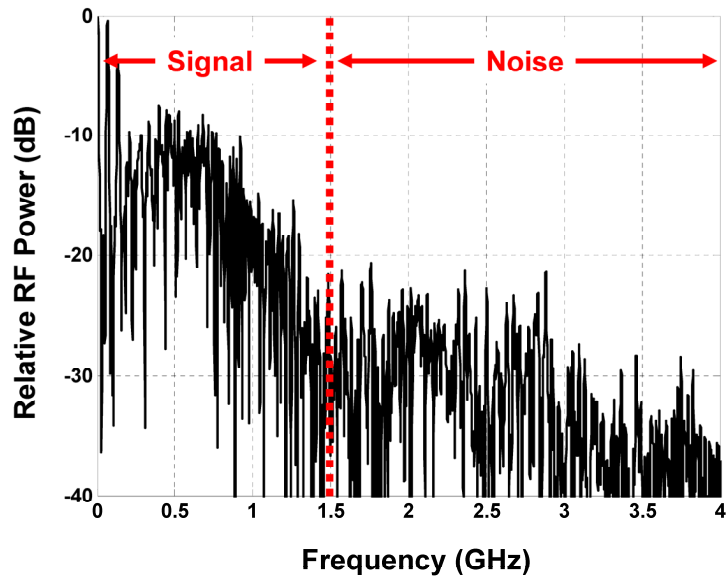

Figure 2: FFT of the encoded time-domain gas signal measured using a 4 GHz oscilloscope. The signal content of the waveform lies primarily below 1.5 GHz. Digital signal processing techniques, such as oversampling, may increase the signal-to-noise ratio in a single shot.

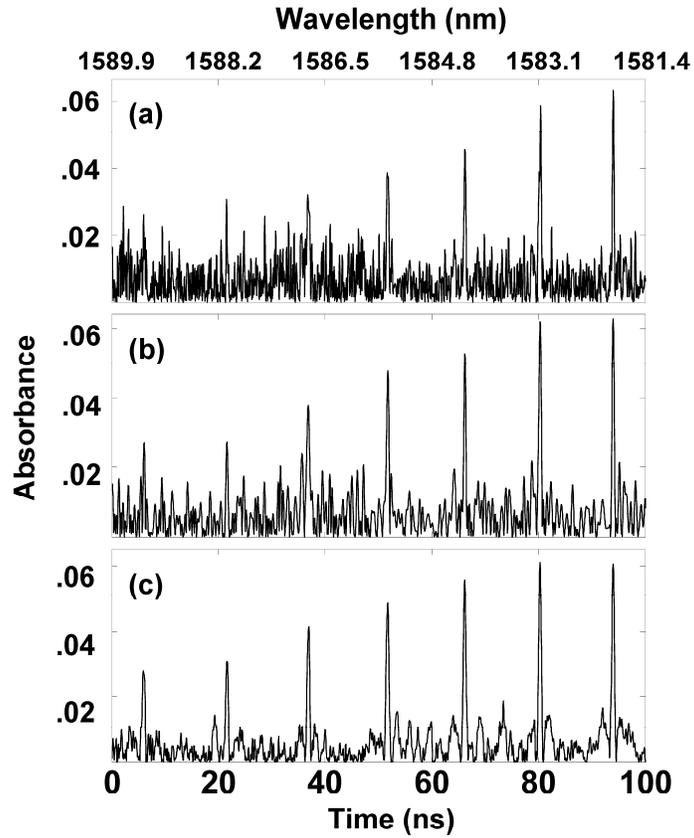

Figure 3: Measured absorption lines of a carbon monoxide sample using CWEETS spectrometer: (a) single shot with high resolution (8 pm or 950 MHz), (b) SNR enhancement by oversampling in a single shot with reduced resolution (47 pm), (c) further SNR enhancement by averaging 100 consecutive pulses with high resolution (8 pm or 950 MHz).

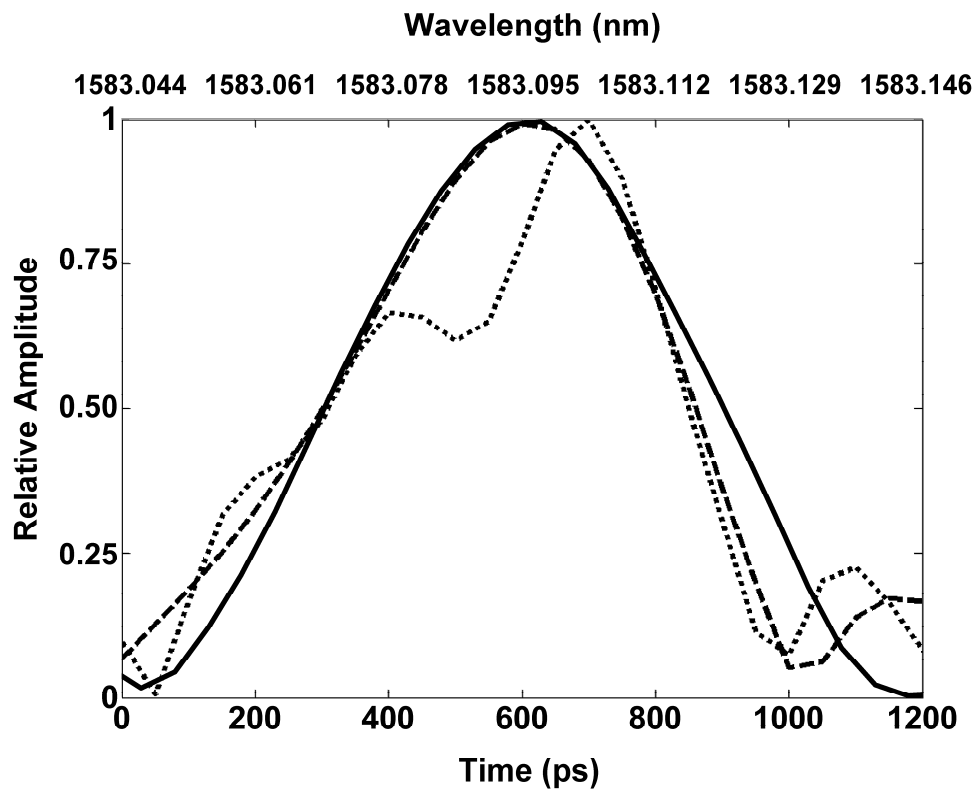

Figure 4: Single absorption-line measurement of carbon monoxide sample showing a line width of approximately 50 pm. Dotted line: single shot with high resolution (8 pm or 950 MHz). Dashed line: SNR enhancement by oversampling in a single shot with reduced resolution (47 pm). Solid line: further SNR enhancement by averaging 100 consecutive pulses with high resolution (8 pm or 950 MHz).